\documentclass{ragtime} 

\usepackage{times}
\usepackage{wrapfig}
\usepackage{lipsum}
\usepackage{tabularx}
\usepackage{txfonts}
\title[Detection of chaotic behavior in time series]{Detection of chaotic behavior in time series }

\author[R. Pánis, M. Kolo\v{s}, Z. Stuchl\'{i}k 
       ]   
       {Radim Pánis\at{1,a} 
       Martin Kolo\v{s}\at[]{1,b} 
                              and  Zden\v{e}k Stuchl\'{i}k\at[]{1,c}           
        \\
        \ins{1}Research Centre for Theoretical Physics and Astrophysics, Institute of Physics, 
        \splitins[1]Silesian University in Opava, Bezru\v{c}ovo n\'am.~13,  CZ-746\,01 Opava,
        Czech Republic\\
               \ins{a}\Email{pan0010@slu.cz}\\
               \ins{b}\Email{martin.kolos@physics.slu.cz} \\
               \ins{c}\Email{zdenek.stuchlik@physics.slu.cz}} 

\newcommand{\Schw}{Schwarzschild}

\def\beq{\begin{equation}}
\def\eeq{\end{equation}}
\def\bea{\begin{eqnarray}}
\def\eea{\end{eqnarray}}

\graphicspath{{pan/}}

\begin{document}

\begin{abstract}

Deterministic chaos is phenomenon from nonlinear dynamics and it belongs to  greatest advances of twentieth-century science. Chaotic behavior appears apart of mathematical equations also in wide range in observable nature, so as in there originating time series. Chaos in time series resembles stochastic behavior, but apart of randomness it is totally deterministic and therefore chaotic data can provide us useful information. Therefore it is essential to have methods, which are able to detect chaos in time series, moreover to distinguish chaotic data from stochastic one. Here we present and discuss the performance of standard and machine learning methods for chaos detection and its implementation on two well known simple chaotic discrete dynamical systems - Logistic map and Tent map, which fit to the most of the definitions of chaos.
\end{abstract}



\section{Introduction}\label{intro}

As already mentioned chaotic behavior is natural phenomena and the first touch with chaos one assigns the  study of the three-body problem by Henri Poincaré  (1880).  Mathematically,  deterministic chaos is the phenomenon which can arise in  dynamical systems.  In  discrete dynamical systems, chaos can appear even for one dimensional systems, while in continuous dynamical systems, deterministic chaos can only arise in three or more dimensions. Finite-dimensional linear systems cannot produce chaos, for a dynamical system to produce chaotic behavior, it must be  either  infinite-dimensional or nonlinear. The main properties of chaotic systems are e.g. the most known sensitivity to initial conditions or more "mainstream" term - butterfly effect, which means that, even the arbitrarily close initial points can evolve into significantly different trajectories.  Other property one should mention  are strange attractors. While most of the types of nice regular systems  provide very simple attractors, such as points or circular curves called limiting cycles, chaotic motion leads to something what is known as the strange attractor, which is attractor with magnificent details and great complexity.

With chaos it is not easy, so we have some dynamical systems, which are chaotic everywhere,  but there are also cases, where chaotic behavior appears only in a subset of the  phase space, then a large set of initial conditions leads to orbits that converge to this chaotic region. Chaotic behavior is for some systems as for example for logistic and tent map easy controllable and it depends on one parameter. Generally,  bifurcation occurs when  small smooth change of "chaotic parameter"  (the bifurcation parameter) makes  abrupt  or topological change in  behavior of the system. This dependency (or period-doubling  transition from some  N-point to an 2N-point attractor) on the chaotic parameter is nicely shown on the bifurcation diagram.
For the purpose of presenting   of the nonlinear  methods of chaos detection we use superposition of graphics, where we display on the bifurcation diagram the estimates of chaotic behavior by later mentioned methods. We applied these methods on time series belonging to given  "chaotic parameter" and we plot this estimation of chaoticity on the bifurcation diagram.

Strange attractors are special and they even have a fractal structure, and therefore one can calculate fractal dimension of them, which can provide useful information about the system even when  the equations of the dynamical system are unknown and we can observe only one of its coordinates. For estimation of fractal dimension we present two methods, namely,  Box-counting and Correlation dimension, classical approach of studying dynamical systems is the calculation of Lyapunov exponent, for  this, one needs to know the rules of evolution for given dynamical system  (equations), we however try to work only with one dimensional input, so we  also use a numerical approach and compare both results. Another approach we use  is recurrence quantification analysis (RQA),  roughly speaking, it is numerical description of recurrence graph - graphical tool for investigating properties of dynamical systems. The last approach is lately again very popular Machine learning  (ML), which has many advantages when dealing with nonlinear data and has proved its abilities in many useful applications. We use for our purpose all seven possible ML implementations build in  Mathematica software, which is  also used for all the calculations.

These methods are also used in \citep{2019EPJC...79..479P}, where the information about the matter dynamics and electromagnetic field structure around  compact object (black hole or neutron star) are provided, while the  chaotic charged test particles dynamics around a \Schw{} black hole immersed in an external uniform magnetic field is examined.

\section{Methods of Detection of chaotic behavior in time series \label{logistic}}


\subsection{Box-counting method \label{boxcount}}

Box-counting ($D_0$) or box dimension is one of the most widely used estimations of fractal dimension. The calculation and empirical estimation of this method is quiet simple compared to another ones. We present the general idea behind the algorithm, for more detailed description one can look at \citep{1991fcpl.book.....S}. For a  set $ S $ in a Euclidean space $ \mathbb{R}^n  $ we define Box-counting as
\beq
D_0 = \lim_{\epsilon \to 0}  \frac{\ln N(\epsilon)}{ \ln \frac{1}{\epsilon}},
\eeq
where $N(\epsilon)$ is the number of boxes of side length  $\epsilon$ required to cover the set. Dimension of $S$ is estimated by seeing how the logarithmic rate of $N(\epsilon)$ increase as ${\epsilon \to 0}$, or in  words as we make the grid finer.

\subsection{Correlation dimension  \label{correlation}}

Very popular tool for detecting chaos in experimental data is calculation of the Correlation dimension ($D_2$). The general idea behind computing correlation dimension is to find out for some small $\epsilon $  the number of points $ C(\epsilon)$ (correlation sum),  which Euclidean distance is smaller than $\epsilon$. 
\beq
D_2 = \lim_{\epsilon \to 0}  \frac{\ln C(\epsilon)}{ \ln \epsilon},
\eeq
one computes this for various number of  $ \epsilon $ and $ D_2 $ can be then approximated again by fitting of the logarithmic values.

It is worth mentioning  that $ D_2 $  and  $ D_0 $  is part of $ D_q  $ family of fractal dimensions \citep{1991fcpl.book.....S}  defined as 
\beq
D_q =    \lim_{\epsilon \to 0} \frac{1}{q - 1}  \frac{\ln \sum_k p ^q _k }{ \ln \epsilon}   \quad   - \infty \leq q \leq \infty,
\eeq
where, $p_k$ denotes relative frequency with which  fractal's points are falling inside the k-th cell.

For $ q = 0 $ we obtain already mentioned Box-counting dimension, for $ q  \to 1 $ we obtain information dimension, which numerator is  denoted as the Shannon's entropy and for  $ q = 2 $ we obtain correlation dimension.

There is several  algorithm approaches of calculation correlation dimension, lets just mention the approximation of $C(\epsilon) $ published in \citep{1983PhRvL..50..346G}:
\beq
\hat{C}(\epsilon)= \lim_{N \to \infty} \frac{2}{N ( N - 1 )} \mathop{\sum_{i < j  }} H ( \epsilon - | x_i - x_j |),
\eeq
where $H$ is Heaviside step function.
 
When using the nonlinear methods one should not omit importance of embedding dimension and then also closely connection to Takens's theorem  about reconstruction of state space from sequence of observations \citep{1981LNM...898..366T}. Embedding dimension creates from series of length $N + m - 1 $ for some given $m$ series of $N$ vectors, where i-th component looks like:
\beq
x_i=(x_{i-m+1}, x_{i-m+2}, ...,x_i) \in \mathbb{R}^m .
\eeq
One of the purposes of embedding dimension, in context of Correlation dimension is to distinguish between chaotic and random time series. Where by chaotic series for  increasing $m \in \mathbb{N}$ fractal dimension estimation stabilizes at some value $D < m$, while for random series, dimension goes along with $m$ to infinity.

\subsection{Lyapunov exponent \label{lyapunov}}

Lyapunov exponents apart of previous methods are originally used for investigation of dynamical systems, or rather said not any fractals. In short, it is a number that describe the amount of separation of  trajectories which are infinitesimally close. Near trajectories in chaotic systems diverge exponentially, what leads to positive Lyapunov exponents. The amount of separation can differ for various directions of initial separation vector. Because of this fact, there is a spectrum of Lyapunov exponents, which corresponds to the phase space dimension.

For example if we consider logistic map $ f(x) = r x(1-x), r \in [0,4]$ as an  typical example of simple chaotic system, which is more precisely one-dimensional nonlinear difference equation. The Lyapunov exponent can be calculated directly from the expression of the  $f$ function from the formula
\beq \label{lyap}
\lambda(r) = \lim_{N \to \infty} \frac{1}{N}  \mathop{\sum_{n = 0 }}_{}^{N-1} \ln [ \, | f'(f^n (x_0)) | \,  ].  
\eeq
However we try to work only with time series inputs which is not allowing us to use such a formula. We assumed in  future to  apply our methods on observational data from the telescope.  This approach leads us to use method, which is determining Lyapunov exponents from time series.
 The  maximal Lyapunov exponent characterizes the spectra and therefore denotes amount of predictability for some dynamical system. It can be calculated without knowledge of a model which produces the time series.
We use the method based on the statistical properties of the  divergence of neighboring trajectories approach introduced by Kantz, \citep{1994PhLA..185...77K}. Our algorithm applied to logistic map is very similar to formula \ref{lyap} which could be  found for example also in \citep{enns2001nonlinear}.

\subsection{Recurrence quantification analysis \label{xRQAx}}

The recurrence quantification analysis is quiet widely used tool for investigating the state space trajectories. Simply said it determines the number and duration of recurrences of a dynamical system. RQA is developed since 1992  \citep{1992PhLA..171..199Z,2008EPJST.164....3M}, where the novel approach based on averaging along with the way of setting the correct input parameters, which provide more accurate RQA measures is presented in \citep{2020arXiv201015085B}. Recurrence plot provides a graphical tool for observing periodicity of phase space trajectories and  was introduced in  \citep{1987EL......4..973E}. This observing is possible through visualization  of a symmetrical square matrix, in which the  elements correspond to times at which a state of a dynamical system recurs.

One can define RP which measures recurrences of a trajectory $ x_i \in  R^d$  in phase space
\beq
R_{i,j}= H(\epsilon - \| x_i - x_j\|  \mid ) \quad  i, j = 1, ...,N,
\eeq
where $N$ is the number of measured points $x_i$, $\epsilon$ is a threshold distance  and $ \| \cdot \|$ is a norm.  From this equation we obtain the already mentioned symmetrical square matrix of zeroes and ones. When we will represent this two repeating elements with different colors in a plot we obtain the discussed RP. Threshold value parameter determinate density of RP plot.

RQA tools with well established short forms, which we use for investigating of chaotic trajectories are:
\begin{enumerate}
\item RR - The recurrence rate is simplest tool,  which measures  density of recurrence points  in the recurrence plot, or in another words, it counts the number of ones in RP and divides them by number of all elements in the matrix. RR reflects the chance that some  state of the system  will recur
 \beq
 RR = \frac{1}{N^2} \sum_{i,j=1}^N R_{i,j}.
 \eeq
\item  DET -   Determinism is  rate of recurrence points which build diagonal lines. DET determines how predictable the system is
\beq
DET  = \frac{\sum_{l=l_{min}}^N l P(l)}{ \sum_{i,j=1}^N R_{i,j}}, 
\eeq
where  $P(l)$ denotes the frequency distribution of lengths $l$  of the diagonal lines.
\item  LL - Is average diagonal line length, which is in relation with the time of predictability  of the system. It reflects the average time for which any two parts of trajectory are close, this time can be denoted as mean prediction time 
\beq
LL = \frac{\sum_{l=l_{min}}^N l P(l)}{\sum_{l=l_{min}}^N  P(l)}.
\eeq
\item  ENTR - Entropy or the Shannon entropy  of the probability distribution of the diagonal line lengths p(l), which are reflecting complexity of the system's deterministic structure
\beq
   ENTR =   - \sum_{l=l_{min}}^N  p(l) \ln p(l),
\eeq
where $p(l)$ is probability  that a diagonal line is exactly of the length $l$  can be estimated from the frequency distribution $P(l)$ with $p(l) = \frac{P(l)}{\sum_{l=l_{min}}^N  P(l)}$.
\end{enumerate}

\subsection{Machine learning  \label{machinelearning}}

Machine learning is very powerful tool, which founds application in many fields, lets just mention language translating algorithms, computer vision, beating best Go player in the world  \citep{2017Natur.550..354S}, or chess programs of different architecture  \citep{2017arXiv171201815S} in an incredible fashion. In this work only basic principles of machine learning are presented.
 Roughly speaking machine learning is field of computer science, strongly connected to another fields  as optimization, statistics, linear algebra, etc.
 
Its beginning goes to 1950's and as many inventions in computer science or better said in science in general. Machine learning was not invented by single person, let's only mention A. Samuel, who used first  the term "Machine learning". Machine learning  is using algorithms on data samples to discover known or unknown patterns in data, this dividing of patterns leads to  basic divisions  of machine learning and namely supervised, semi-supervised and unsupervised learning.
The wide range of applications announces the good ability of handling nonlinear data.
Our intention of using machine learning is to decide whether a trajectory of a particle is chaotic or not. For this purpose we use supervised machine learning, where we train various ML algorithms  with samples calculated by classical methods already described, namely, (Box-counting, Correlation dimension, Lyapunov exponent, RQA - RR, DET, LL, ENTR ) with effort to use all the different properties of them and the training set consists overall of 100 examples. 

\begin{figure*}
\includegraphics[width=\hsize]{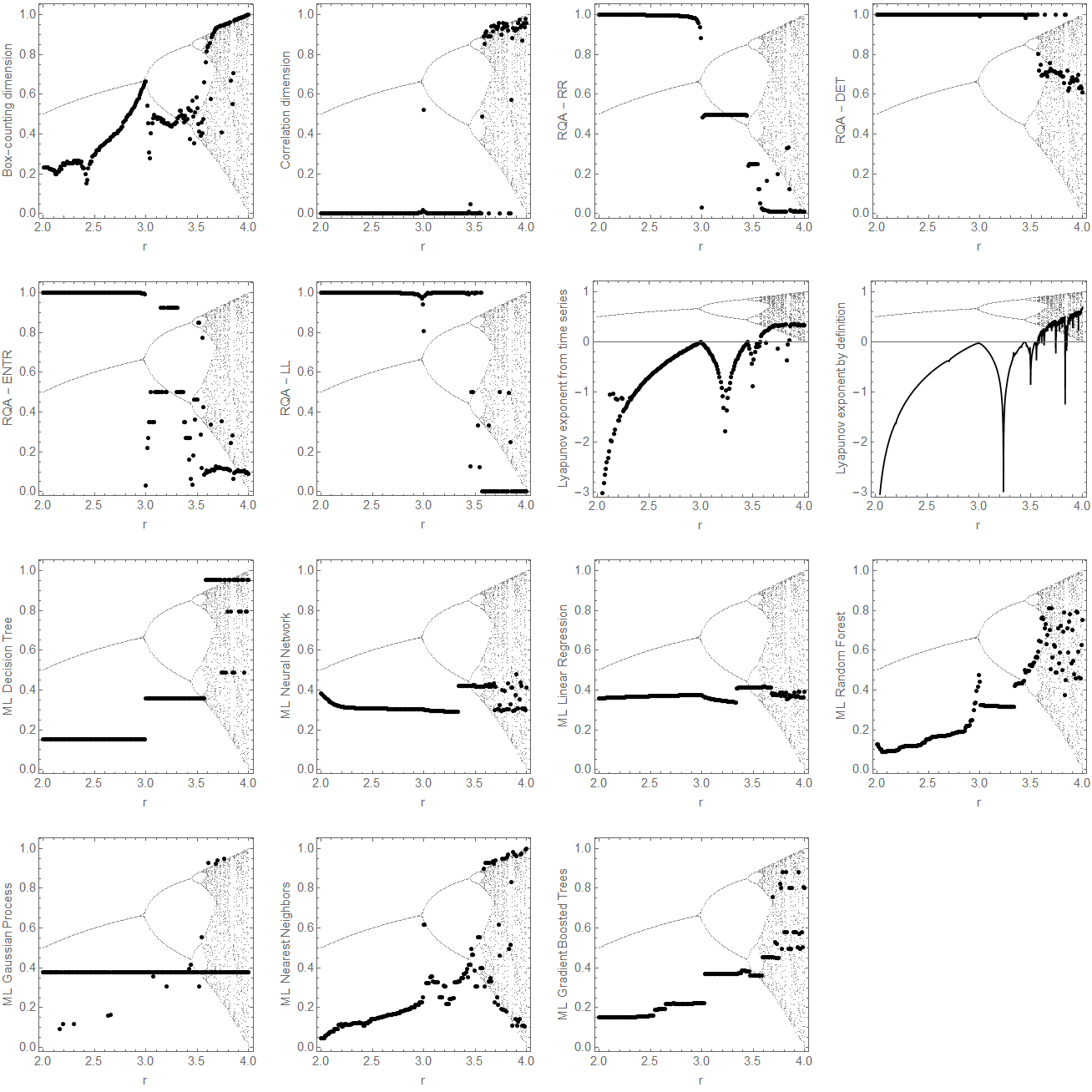}

\caption{\label{porovnanie}
Comparison of different nonlinear methods for time series generated by logistic map $ x_{n+1} = r x_{n}(1-x_n)$. For different parameter $r \in [2,4] $ we generate 10000 points series with initial value  $x_0=0.1$.  All the methods detect more chaoticity when we enter the chaotic region for value $r >r_0\approx~3.56995$. RQA tools however, are working in reverse fashion as one can see, they denote more chaotic regions with lower numbers and vice versa, the data (chaoticity estimations) for RQA-LL and RQA-ENTR  have been transformed into interval $[0,1]$.
}
\end{figure*}


\begin{figure*}
\includegraphics[width=\hsize]{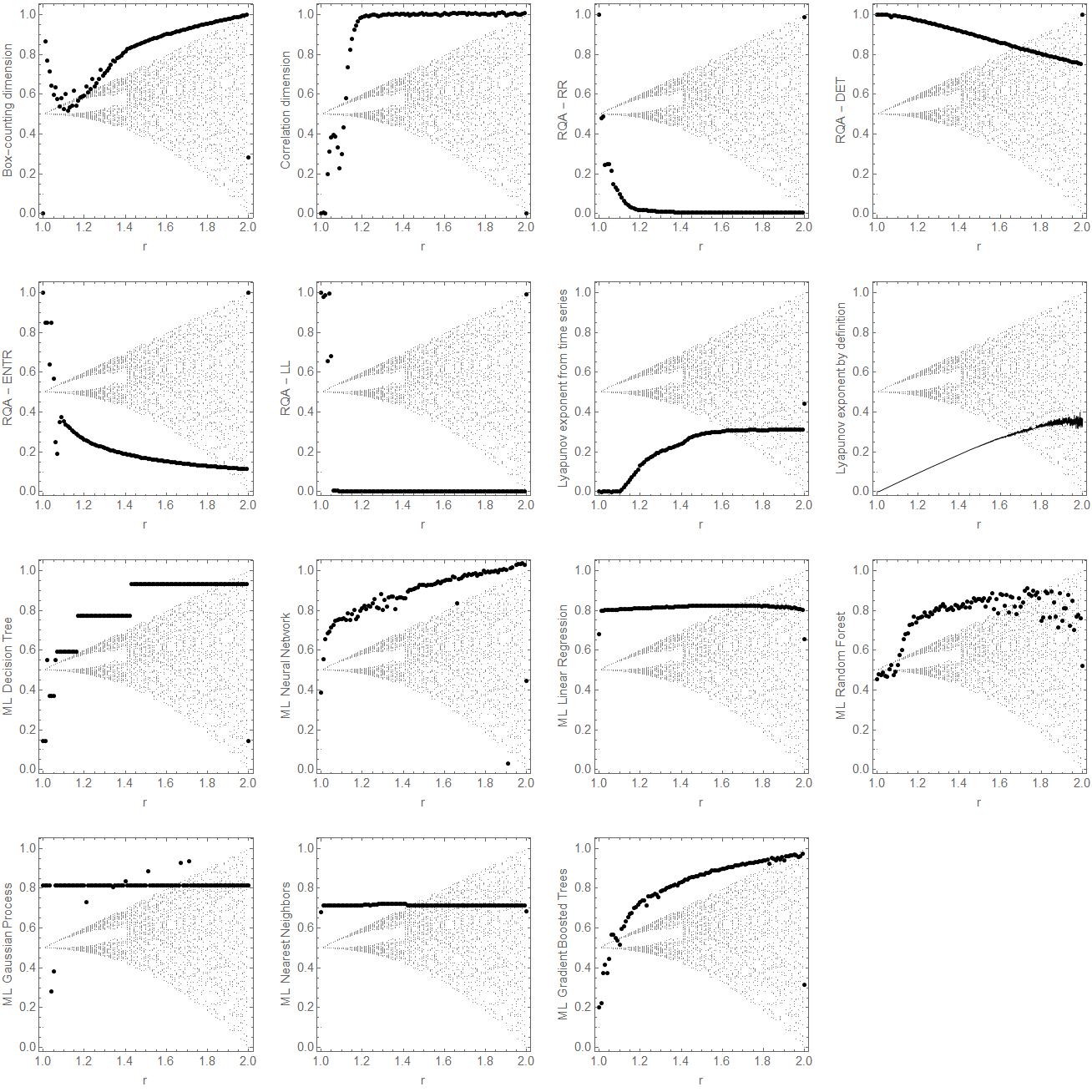}

\caption{\label{porovnanie2}
Comparison of different nonlinear methods for time series generated by tent map defined as
$x_{{n+1}}=f_{r }(x_{n})={\begin{cases}r x_{n}&{\mathrm  {for}}~~x_{n}<{\frac  {1}{2}}\\\\r (1-x_{n})&{\mathrm  {for}}~~{ x_{n}}\geq \frac  {1}{2}\end{cases}}$. For different parameter $r \in [1,2]$ we generate 10000 points series with initial value  $x_0=0.1$.
}
\end{figure*}


\begin{table}
\caption{\label{tab2} Comparison of 
time in seconds required for different nonlinear methods applied on  time series generated by logistic map. For  parameter $r$ varying from $ 2$ to $4$ with the step of $0.01$ leads to $200$ time series, with initial value  $x_0=0.1$ we did set up the iterations to length  $ 100, 1000$ and $ 10000 $ and compared the time needed for the calculation  for given methods.}
\begin{center}
    \begin{tabular}{| l | l | l | l |}
    \hline
    Method / Length & 100 & 1000 & 10000\\ \hline
   Box-count & 0.354 & 0.416 & 3.519  \\ \hline
   Correlation dim. & 4.155& 26.54 & 2353 \\ \hline
    Lyapunov exp. & 3.294 & 119.5 & 12574 \\    \hline
  RQA- RR & 0.164 & 6.332 & 507.8 \\    \hline
   RQA-DET & 0.209 & 6.069 &  916.9 \\    \hline
   RQA-ENTR &  0.197 &  7.247 &  1128  \\    \hline
   RQA-LAM & 0.197 & 7.248 & 1128 \\    \hline
       ML-Rand. For. & 0.677 & 1.212 & 7.479 \\    \hline
    \end{tabular}
\end{center}
\end{table}

\section{Conclusion}\label{conc}

Deterministic chaos is hardly predictable and apparently random behavior which can appear in dynamical systems. Classical example of such nonlinear dynamical system is logistic map, denoted by quadratic recurrence equation
\beq
 x_{n+1} = r x_{n}(1-x_n). \label{logistic}
\eeq
Given the initial value $x_0\in(0,1)$, the logistic map (\ref{logistic}) will generate sequence of real numbers $x_n\in(0,1)$. The behavior of such sequence $x_n$ strongly depends on logistic map parameter $r\in[2,4]$. Roughly speaking is the behavior on the interval $r\in(2,r_0)$  regular (rather predictable) and on the interval $r\in[r_0,4]$ is chaotic (hardly predictable or rather unpredictable) with some occasional "islands of regularity". The transition between regular and chaotic behaviors happens for parameter $r=r_0\approx~3.56995$. 
Bifurcation diagram for logistic map, with asymptotically approached values of the sequence, is shown in Fig. \ref{porovnanie} depicted by gray background points.


For chaos detection and chaotic behavior description in sequences of numbers following methods have been tested: Box-counting method (section \ref{boxcount}), Correlation dimension (\ref{correlation}), Lyapunov exponent (\ref{lyapunov}), RQA (\ref{xRQAx}) and diferent  machine learning algorithms (\ref{machinelearning}). All of the algorithms we use in this article are capable to use  one dimensional sequence of real numbers  (time series) as input. From the point of observation is important to distinguish between chaotic sequences ruled by some (unknown) laws  and random sequences obeying for example some stochastic distribution, which can contribute to the noise part in the detected signal. The theoretical boundary of distinguishing between chaos and random sequences is sequence length \citep{Ott:1993:book:}. If the nonlinear dynamical system has many degrees of freedom, then for short sequences is impossible to tell.


To clear up how the nonlinear methods works is shown on four representative sequences $ x_n^{(1)}, x_n^{(2)}, x_n^{(3)}, x_n^{(4)}$ of length $10^4$, where $ x_n^{(1)}$ is regular sequence, $ x_n^{(2)}$ is weakly chaotic sequence generated by logistic map with $r = 3.6 $, $x_n^{(3)}$ is strongly chaotic sequence generated by logistic map with $ r = 4 $, and $x_n^{(4)}$ is sequence of pseudo-random numbers. The results are presented in Tab. \ref{tab1}.

In detail we have tested the nonlinear methods on sequences of numbers generated by logistic and tent map for various values of parameter $r$, see Fig. \ref{porovnanie} and  \ref{porovnanie2}, the application/testing of the methods on a specific task can be found in  \citep{2019EPJC...79..479P}.
 The important common sign of all these methods is that, they are able to detect more chaoticity when divergence of trajectories in bifurcation diagram occurs, in other words, when there is more than one fixed point.  This fact we can observe for example when $r=3$ for logistic map in Fig. \ref{porovnanie} and for $r$  little above value $1$  for tent map in Fig. \ref{porovnanie2}. However, this is not undeniable true for all the cases, what leads to different estimations of chaoticity for different methods. Box-counting and Lyapunov exponent method show approximately linear behavior when moving to point, where the divergence of the trajectories begins, while when moving to such a point chaoticity is increasing and decreasing by leaving it. Correlation dimension shows little bit different behavior, when nonchaotic region is denoted by values close to zero and the level of chaos starts rapidly grow only from small distance from  the region of divergence of trajectories in bifurcation diagram, this behavior is also observable by RQA tools, however in reversed fashion. Satisfying is the detection of chaoticity for logistic map for the $r \geq 3.56995 $, when all these methods are showing the highest values of chaoticity,  in this region are also "islands of regularity" though, what means not all the estimations of chaos in this region should be high. By tent map is the behavior of the methods also satisfying, we should not omit  the fact the divergence of trajectories starts from very small values above $r=1$ and grows with $r$ moving to value $r=2$, where from our definition of tent map and initial condition value $x_0=0.1$ the behavior gets suddenly quiet regular again.

As expected different machine learning algorithms produce different results. Random Forest algorithm seems to produce quiet reliable result -  our estimation of chaoticity is getting higher most of the time as the  $r$ parameter grows, the exceptions could be explained by the fact there are also "islands of regularity" in the chaotic region.

The time spent for calculation of the individual graphics in Fig. \ref{porovnanie}
is presented in Tab. \ref{tab2}, where we in addition present also the times consumed in the cases for shorter time series in order to catch the nonlinear growth of  time consumption with larger input. From here we can see that the Box-counting method is the fastest, but when looking on the results the other methods seem to be more precise. The time spent also strongly depends on the sequence length - such dependence is nonlinear in the case of almost all the methods. For most of the methods we have programmed and tested several variants for each  of them according to algorithm architecture and settings of adjustable parameters. The goal was to obtain good and also  computing   time acceptable results,  the code in Mathematica is available on the GitHub repositary at \citep{Git}.

In previous texts and presented Figures \ref{porovnanie}, \ref{porovnanie2}, and tables \ref{tab1} and \ref{tab2} we provide brief overview about the common and novel methods used in nonlinear time series analysis, which can provide useful hints for potential users in the sense of finding suitable method when studying the nonlinear phenomena.
    Proclaim any here discussed method as the best would be misleading, and this  ambiguity   expresses aptly  the saying, that every complicated question has a simple answer which is wrong. When choosing a method for particular purpose, one should have a good idea about the data structure  as well as have in mind the definitions of the methods when discussing the results in the context of the underlying physics. Considering the main properties of data, when choosing a method, one should definitely have in mind the available hardware capabilities along with the dimensions of the data. 

For very large data sets definitely the Box-counting and ML methods should be suitable, however
when considering ML, the available data for training set constriction have to be mentioned.
Lyapunov exponents are popular in physics from the point of interpretation and connection to the direct physical properties of the systems. RQA in the trade-off with  longer computational time needed provide more numerical descriptions of the data, the connection to the physical properties in terms of Blazars can be found in \citep{2020arXiv201015085B} and the enhancing of the computational time needed could be the point of future research in a context of GPU usage.

\begin{table}
\caption{\label{tab1} 
Comparison of values of different nonlinear methods applied on regular, chaotic and pseudo-random generated series of the length 10 000. }
\begin{center}
    \begin{tabular}{| l | l | l | l | l |}
    \hline
    Met./ Type & Regular & Weak ch. & Strong ch. & P.-rand.\\ \hline
   Box-count & 0.0969 & 0.8343 & 1   & 1 \\ \hline
   Corr. dim. & $5*10^{-15}$ & 0.8943 & 0.8735 & 0.9978  \\ \hline
    Lyap. exp. &  $-7*10^{-16}$ & 0.1889 & 0.3289  &   0.0842  \\    \hline
  RQA-RR & 0.4999 & 0.0183 & 0.0116 & 0.0057 \\    \hline
   RQA-DET & 1 & 0.6959 & 0.6114   &  0.0113 \\    \hline
   RQA-ENTR & 8.5152  & 0.8951 & 0.8202   &   0.0296 \\    \hline
   RQA-LAM & 5001 & 5.6299 & 2.8507  &  2.0006 \\    \hline
     ML-Rand. For. & 0.5915 & 0.6917 & 0.7869  & 0.3788  \\    \hline
    \end{tabular}
\end{center}
\end{table}

\section*{Acknowledgments}

This work was supported by the Student Grant Foundation of the Silesian University in Opava, Grant No. $\mathrm{SGF/4/2020}$, which has been carried out within the EU OPSRE project entitled ``Improving the quality of the internal grant scheme of the Silesian University in Opava'', reg. number: $\mathrm{CZ.02.2.69/0.0/0.0/19\_073/0016951}$. ZS thanks for the support of the Institute of Physics and Research Centre of Theoretical Physics and Astrophysics, at the Silesian University in Opava. RP acknowledge the institutional support of Silesian University in Opava and the grant SGS/12/2019. We are very thankful to the anonymous referee for his/her thorough and careful reading of the paper and very useful comments and suggestions which helped improve the presentation of the paper significantly.




\def\prc{Phys. Rev. C}
\def\pre{Phys. Rev. E}
\def\prd{Phys. Rev. D}
\def\jcap{Journal of Cosmology and Astroparticle Physics}
\def\apss{Astrophysics and Space Science}
\def\mnras{Monthly Notices of the Royal Astronomical Society}
\def\apj{The Astrophysical Journal}
\def\aap{Astronomy and Astrophysics}
\def\actaa{Acta Astronomica}
\def\pasj{Publications of the Astronomical Society of Japan}
\def\apjl{Astrophysical Journal Letters}
\def\pasa{Publications Astronomical Society of Australia}
\def\nat{Nature}
\def\physrep{Physics Reports}
\def\araa{Annual Review of Astronomy and Astrophysics}
\def\apjs{The Astrophysical Journal Supplement}
\def\aapr{The Astronomy and Astrophysics Review}
\def\procspie{Proceedings of the SPIE}

\def\pasp{PASP}
\def\aj{AJ}
\def\nat{Nature}
\def\nar{NewAR}
\def\na{NewA}
\def\icarus{Icar}
\def\araa{ARA\&A}
\def\aplett{Astrophysical Letters}
\def\prl{Physical Review Letters}

\def\prc{Phys. Rev. C}
\def\pre{Phys. Rev. E}
\def\prd{Phys. Rev. D}
\def\jcap{Journal of Cosmology and Astroparticle Physics}
\def\apss{Astrophysics and Space Science}
\def\mnras{Mon. Not. R. Astron Soc.}
\def\apj{The Astrophysical Journal}
\def\aap{Astron. Astrophys.}
\def\actaa{Acta Astronomica}
\def\pasj{Publications of the Astronomical Society of Japan}
\def\apjl{Astrophysical Journal Letters}
\def\pasa{Publications Astronomical Society of Australia}
\def\nat{Nature}
\def\physrep{Phys. Rep.}
\def\araa{Annu. Rev. Astron. Astrophys.}
\def\apjs{The Astrophysical Journal Supplement}
\def\aapr{The Astronomy and Astrophysics Review}

\def\mdash{---}
\bibliography{pan.bib}

\begin{thebibliography}{14}
\expandafter\ifx\csname natexlab\endcsname\relax\def\natexlab#1{#1}\fi
\expandafter\ifx\csname url\endcsname\relax
  \def\url#1{\texttt{#1}}\fi
\expandafter\ifx\csname urlprefix\endcsname\relax\def\urlprefix{URL }\fi
\providecommand{\selectlanguage}[1]{\relax}
\providecommand{\eprint}[2][]{\url{#2}}

\bibitem[{{Bhatta} et~al.(2020){Bhatta}, {P{\'a}nis} and
  {Stuchl{\'\i}k}}]{2020arXiv201015085B}
{Bhatta}, G., {P{\'a}nis}, R. and {Stuchl{\'\i}k}, Z. (2020), {Deterministic
  Aspect of the $\gamma$-ray Variability in Blazars}, \emph{arXiv e-prints},
  arXiv:2010.15085, \eprint{2010.15085}.

\bibitem[{{Eckmann} et~al.(1987){Eckmann}, {Oliffson Kamphorst} and
  {Ruelle}}]{1987EL......4..973E}
{Eckmann}, J.-P., {Oliffson Kamphorst}, S. and {Ruelle}, D. (1987), {Recurrence
  plots of dynamical systems}, \emph{EPL (Europhysics Letters)}, \textbf{4}, p.
  973.

\bibitem[{Enns(2001)}]{enns2001nonlinear}
Enns, R. (2001), \emph{Nonlinear physics with Mathematica for scientists and
  engineers}, Birkhauser, Boston, ISBN 978-0-8176-4223-5.

\bibitem[{{Grassberger} and {Procaccia}(1983)}]{1983PhRvL..50..346G}
{Grassberger}, P. and {Procaccia}, I. (1983), {Characterization of strange
  attractors}, \emph{Physical Review Letters}, \textbf{50}, pp. 346--349.

\bibitem[{{Kantz}(1994)}]{1994PhLA..185...77K}
{Kantz}, H. (1994), {A robust method to estimate the maximal Lyapunov exponent
  of a time series}, \emph{Physics Letters A}, \textbf{185}, pp. 77--87.

\bibitem[{{Marwan}(2008)}]{2008EPJST.164....3M}
{Marwan}, N. (2008), {A historical review of recurrence plots}, \emph{European
  Physical Journal Special Topics}, \textbf{164}, pp. 3--12,
  \eprint{1709.09971}.

\bibitem[{{Ott}(1993)}]{Ott:1993:book:}
{Ott}, E. (1993), \emph{{Chaos in dynamical systems}}, Cambridge University
  Press.

\bibitem[{{P{\'a}nis} et~al.(2019){P{\'a}nis}, {Kolo{\v{s}}} and
  {Stuchl{\'\i}k}}]{2019EPJC...79..479P}
{P{\'a}nis}, R., {Kolo{\v{s}}}, M. and {Stuchl{\'\i}k}, Z. (2019),
  {Determination of chaotic behaviour in time series generated by charged
  particle motion around magnetized Schwarzschild black holes}, \emph{European
  Physical Journal C}, \textbf{79}(6), 479, \eprint{1905.01186}.

\bibitem[{Pánis(2019)}]{Git}
Pánis, R. (2019), Chaos detection,
  \url{https://github.com/radim525/Chaos-detection}.

\bibitem[{{Schroeder}(1991)}]{1991fcpl.book.....S}
{Schroeder}, M. (1991), \emph{{Fractals, chaos, power laws. Minutes from an
  infinte paradise}}.

\bibitem[{{Silver} et~al.(2017{\natexlab{a}}){Silver}, {Hubert},
  {Schrittwieser}, {Antonoglou}, {Lai}, {Guez}, {Lanctot}, {Sifre}, {Kumaran},
  {Graepel}, {Lillicrap}, {Simonyan} and {Hassabis}}]{2017arXiv171201815S}
{Silver}, D., {Hubert}, T., {Schrittwieser}, J., {Antonoglou}, I., {Lai}, M.,
  {Guez}, A., {Lanctot}, M., {Sifre}, L., {Kumaran}, D., {Graepel}, T.,
  {Lillicrap}, T., {Simonyan}, K. and {Hassabis}, D. (2017{\natexlab{a}}),
  {Mastering Chess and Shogi by Self-Play with a General Reinforcement Learning
  Algorithm}, \emph{ArXiv e-prints}, \eprint{1712.01815}.

\bibitem[{{Silver} et~al.(2017{\natexlab{b}}){Silver}, {Schrittwieser},
  {Simonyan}, {Antonoglou}, {Huang}, {Guez}, {Hubert}, {Baker}, {Lai},
  {Bolton}, {Chen}, {Lillicrap}, {Hui}, {Sifre}, {van den Driessche}, {Graepel}
  and {Hassabis}}]{2017Natur.550..354S}
{Silver}, D., {Schrittwieser}, J., {Simonyan}, K., {Antonoglou}, I., {Huang},
  A., {Guez}, A., {Hubert}, T., {Baker}, L., {Lai}, M., {Bolton}, A., {Chen},
  Y., {Lillicrap}, T., {Hui}, F., {Sifre}, L., {van den Driessche}, G.,
  {Graepel}, T. and {Hassabis}, D. (2017{\natexlab{b}}), {Mastering the game of
  Go without human knowledge}, \emph{\nat}, \textbf{550}, pp. 354--359.

\bibitem[{{Takens}(1981)}]{1981LNM...898..366T}
{Takens}, F. (1981), \emph{{Detecting strange attractors in turbulence}},
  volume 898, p. 366.

\bibitem[{{Zbilut} and {Webber}(1992)}]{1992PhLA..171..199Z}
{Zbilut}, J.~P. and {Webber}, C.~L. (1992), {Embeddings and delays as derived
  from quantification of recurrence plots}, \emph{Physics Letters A},
  \textbf{171}, pp. 199--203.

\end{thebibliography}

\end{document}